\documentclass[prd,superscriptaddress,showpacs,amssymb,amsmath,amsfonts,aps,altaffilletter,nofootinbib,letterpaper,twocolumn]{revtex4}

\usepackage{color}
\usepackage{graphicx}
\usepackage{fancyhdr}
\usepackage{float}
\usepackage{ulem}
\usepackage[raggedright]{subfigure}
\usepackage{acronym}
\usepackage{multirow}
\usepackage{array}
\usepackage{hyperref}

\makeatletter
\makeatletter
\def\@fnsymbol#1{\ifcase#1\or * \or  $+$ \or  \$ \or \#  \or \dag \or \ddag \or
$\mathsection$ \or $ \mathparagraph$ \or $\|$  \or \textordfeminine \or \textbul
let   
\or ** \or $++$ \or  \$\$ \or \#\#  \or \dag\dag \or \ddag\ddag \or
$\mathsection\mathsection$ \or $ \mathparagraph\mathparagraph$ \or $\|\|$  \or 
\textordfeminine\textordfeminine \or \textbullet \textbullet \or *** \or $+++$ 
\or  \$\$\$ \or \#\#  \or \dag\dag \or \ddag\ddag \or
$\mathsection \mathsection\mathsection$ \or $ \mathparagraph 
\mathparagraph\mathparagraph$ \or $\|\|\|$  \or 
\textordfeminine\textordfeminine\textordfeminine \or 
\textbullet\textbullet\textbullet \or \else \@ctrerr\fi}
\makeatother

\newcommand\fake[1]{\textcolor{red}{#1}}

\def\thercsid{\relax}
\def\rcsid#1{\def\next##1#1{\def\thercsid{##1}}\next}

\rcsid$Id: s6-cbc-lowmass.tex,v 1.86 2012/01/17 15:59:19 cdcapano Exp $

\renewcommand{\today}{\number\day\space\ifcase\month\or
  January\or February\or March\or April\or May\or June\or
  July\or August\or September\or October\or November\or December\fi
  \space\number\year}

\def\Msun{\ensuremath{\mathrm{M_{\odot}}}}

\def\firstFAR{\ensuremath{\mathrm{1.2\,yr^{-1}}}}
\def\secondFAR{\ensuremath{\mathrm{2.2\,yr^{-1}}}}
\def\thirdFAR{\ensuremath{\mathrm{5.6\,yr^{-1}}}}
\def\expectedLoudestFAR{\ensuremath{\mathrm{2\pm2\,yr^{-1}}}}

\def\HLltpreVeto{\ensuremath{0.21}\,yr}
\def\HVltpreVeto{\ensuremath{0.13}\,yr}
\def\LVltpreVeto{\ensuremath{0.08}\,yr}
\def\HLVltpreVeto{\ensuremath{0.14}\,yr}
\def\totalpreVeto{\ensuremath{0.56}\,yr}
\def\HLlt{\ensuremath{0.17}\,yr}
\def\HVlt{\ensuremath{0.10}\,yr}
\def\LVlt{\ensuremath{0.07}\,yr}
\def\HLVlt{\ensuremath{0.09}\,yr}
\def\totalTime{\ensuremath{0.43}\,yr}

\def\12to18{Abbott:2009qj}
\def\sfive1yr{Collaboration:2009tt}
\def\sfivelvc{S5LowMassLV}

\def\BNShd{\ensuremath{40}}
\def\BNSul{\ensuremath{1.3 \times 10^{-4}}}

\def\NSBHhd{\ensuremath{80}}
\def\NSBHul{\ensuremath{3.1 \times 10^{-5}}}
\def\sNSBHul{\ensuremath{3.6 \times 10^{-5}}}

\def\BBHhd{\ensuremath{90}}
\def\BBHul{\ensuremath{6.4 \times 10^{-6}}}
\def\sBBHul{\ensuremath{7.4 \times 10^{-6}}}

\newcommand\perMpcyr{\ensuremath{\mathrm{Mpc}^{-3} \mathrm{yr}^{-1}}}
\def\dogDate{16 September 2010}

\def\recoveredDogTime{06:42:23 UTC}
\def\dogTrialFAR{1 in $4 \times 10^{4}$ years}
\def\dogFAR{1 in $7,000$ years}

\def\dogFAP{\ensuremath{7 \times 10^{-5}}}

\def\dogSNR{\ensuremath{\rho_c = 12.5}}

\begin{document}

\title{Search for Gravitational Waves from Low Mass Compact Binary Coalescence
in LIGO's Sixth Science Run and Virgo's Science Runs 2 and 3\\ 
LIGO Document P1100034-v18}





\affiliation{LIGO - California Institute of Technology, Pasadena, CA  91125, USA$^\ast$}
\affiliation{California State University Fullerton, Fullerton CA 92831 USA$^\ast$}
\affiliation{SUPA, University of Glasgow, Glasgow, G12 8QQ, United Kingdom$^\ast$}
\affiliation{Laboratoire d'Annecy-le-Vieux de Physique des Particules (LAPP), Universit\'e de Savoie, CNRS/IN2P3, F-74941 Annecy-Le-Vieux, France$^\dagger$}
\affiliation{INFN, Sezione di Napoli $^a$; Universit\`a di Napoli 'Federico II'$^b$ Complesso Universitario di Monte S.Angelo, I-80126 Napoli; Universit\`a di Salerno, Fisciano, I-84084 Salerno$^c$, Italy$^\dagger$}
\affiliation{LIGO - Livingston Observatory, Livingston, LA  70754, USA$^\ast$}
\affiliation{Albert-Einstein-Institut, Max-Planck-Institut f\"ur Gravitationsphysik, D-30167 Hannover, Germany$^\ast$}
\affiliation{Leibniz Universit\"at Hannover, D-30167 Hannover, Germany$^\ast$}
\affiliation{Nikhef, Science Park, Amsterdam, the Netherlands$^a$; VU University Amsterdam, De Boelelaan 1081, 1081 HV Amsterdam, the Netherlands$^b$$^\dagger$}
\affiliation{University of Wisconsin--Milwaukee, Milwaukee, WI  53201, USA$^\ast$}
\affiliation{Stanford University, Stanford, CA  94305, USA$^\ast$}
\affiliation{University of Florida, Gainesville, FL  32611, USA$^\ast$}
\affiliation{Louisiana State University, Baton Rouge, LA  70803, USA$^\ast$}
\affiliation{University of Birmingham, Birmingham, B15 2TT, United Kingdom$^\ast$}
\affiliation{INFN, Sezione di Roma$^a$; Universit\`a 'La Sapienza'$^b$, I-00185 Roma, Italy$^\dagger$}
\affiliation{LIGO - Hanford Observatory, Richland, WA  99352, USA$^\ast$}
\affiliation{Albert-Einstein-Institut, Max-Planck-Institut f\"ur Gravitationsphysik, D-14476 Golm, Germany$^\ast$}
\affiliation{Montana State University, Bozeman, MT 59717, USA$^\ast$}
\affiliation{European Gravitational Observatory (EGO), I-56021 Cascina (PI), Italy$^\dagger$}
\affiliation{Syracuse University, Syracuse, NY  13244, USA$^\ast$}
\affiliation{University of Western Australia, Crawley, WA 6009, Australia$^\ast$}
\affiliation{LIGO - Massachusetts Institute of Technology, Cambridge, MA 02139, USA$^\ast$}
\affiliation{Laboratoire AstroParticule et Cosmologie (APC) Universit\'e Paris Diderot, CNRS: IN2P3, CEA: DSM/IRFU, Observatoire de Paris, 10 rue A.Domon et L.Duquet, 75013 Paris - France$^\dagger$}
\affiliation{Columbia University, New York, NY  10027, USA$^\ast$}
\affiliation{INFN, Sezione di Pisa$^a$; Universit\`a di Pisa$^b$; I-56127 Pisa; Universit\`a di Siena, I-53100 Siena$^c$, Italy$^\dagger$}
\affiliation{The University of Texas at Brownsville and Texas Southmost College, Brownsville, TX  78520, USA$^\ast$}
\affiliation{San Jose State University, San Jose, CA 95192, USA$^\ast$}
\affiliation{Moscow State University, Moscow, 119992, Russia$^\ast$}
\affiliation{LAL, Universit\'e Paris-Sud, IN2P3/CNRS, F-91898 Orsay$^a$; ESPCI, CNRS,  F-75005 Paris$^b$, France$^\dagger$}
\affiliation{NASA/Goddard Space Flight Center, Greenbelt, MD  20771, USA$^\ast$}
\affiliation{The Pennsylvania State University, University Park, PA  16802, USA$^\ast$}
\affiliation{Universit\'e Nice-Sophia-Antipolis, CNRS, Observatoire de la C\^ote d'Azur, F-06304 Nice$^a$; Institut de Physique de Rennes, CNRS, Universit\'e de Rennes 1, 35042 Rennes$^b$, France$^\dagger$}
\affiliation{Laboratoire des Mat\'eriaux Avanc\'es (LMA), IN2P3/CNRS, F-69622 Villeurbanne, Lyon, France$^\dagger$}
\affiliation{Washington State University, Pullman, WA 99164, USA$^\ast$}
\affiliation{INFN, Sezione di Perugia$^a$; Universit\`a di Perugia$^b$, I-06123 Perugia,Italy$^\dagger$}
\affiliation{INFN, Sezione di Firenze, I-50019 Sesto Fiorentino$^a$; Universit\`a degli Studi di Urbino 'Carlo Bo', I-61029 Urbino$^b$, Italy$^\dagger$}
\affiliation{University of Oregon, Eugene, OR  97403, USA$^\ast$}
\affiliation{Laboratoire Kastler Brossel, ENS, CNRS, UPMC, Universit\'e Pierre et Marie Curie, 4 Place Jussieu, F-75005 Paris, France$^\dagger$}
\affiliation{Rutherford Appleton Laboratory, HSIC, Chilton, Didcot, Oxon OX11 0QX United Kingdom$^\ast$}
\affiliation{IM-PAN 00-956 Warsaw$^a$; Astronomical Observatory Warsaw University 00-478 Warsaw$^b$; CAMK-PAN 00-716 Warsaw$^c$; Bia{\l}ystok University 15-424 Bia{\l}ystok$^d$; IPJ 05-400 \'Swierk-Otwock$^e$; Institute of Astronomy 65-265 Zielona G\'ora$^f$,  Poland$^\dagger$}
\affiliation{University of Maryland, College Park, MD 20742 USA$^\ast$}
\affiliation{University of Massachusetts - Amherst, Amherst, MA 01003, USA$^\ast$}
\affiliation{Canadian Institute for Theoretical Astrophysics, University of Toronto, Toronto, Ontario, M5S 3H8, Canada$^\ast$}
\affiliation{Tsinghua University, Beijing 100084 China$^\ast$}
\affiliation{University of Michigan, Ann Arbor, MI  48109, USA$^\ast$}
\affiliation{The University of Mississippi, University, MS 38677, USA$^\ast$}
\affiliation{Charles Sturt University, Wagga Wagga, NSW 2678, Australia$^\ast$}
\affiliation{Caltech-CaRT, Pasadena, CA  91125, USA$^\ast$}
\affiliation{INFN, Sezione di Genova;  I-16146  Genova, Italy$^\dagger$}
\affiliation{Pusan National University, Busan 609-735, Korea$^\ast$}
\affiliation{Carleton College, Northfield, MN  55057, USA$^\ast$}
\affiliation{Australian National University, Canberra, ACT 0200, Australia$^\ast$}
\affiliation{The University of Melbourne, Parkville, VIC 3010, Australia$^\ast$}
\affiliation{Cardiff University, Cardiff, CF24 3AA, United Kingdom$^\ast$}
\affiliation{INFN, Sezione di Roma Tor Vergata$^a$; Universit\`a di Roma Tor Vergata, I-00133 Roma$^b$; Universit\`a dell'Aquila, I-67100 L'Aquila$^c$, Italy$^\dagger$}
\affiliation{University of Salerno, I-84084 Fisciano (Salerno), Italy and INFN (Sezione di Napoli), Italy$^\dagger$}
\affiliation{The University of Sheffield, Sheffield S10 2TN, United Kingdom$^\ast$}
\affiliation{RMKI, H-1121 Budapest, Konkoly Thege Mikl\'os \'ut 29-33, Hungary$^\ast$}
\affiliation{INFN, Gruppo Collegato di Trento$^a$ and Universit\`a di Trento$^b$,  I-38050 Povo, Trento, Italy;   INFN, Sezione di Padova$^c$ and Universit\`a di Padova$^d$, I-35131 Padova, Italy$^\dagger$}
\affiliation{Inter-University Centre for Astronomy and Astrophysics, Pune - 411007, India$^\ast$}
\affiliation{University of Minnesota, Minneapolis, MN 55455, USA$^\ast$}
\affiliation{California Institute of Technology, Pasadena, CA  91125, USA$^\ast$}
\affiliation{Northwestern University, Evanston, IL  60208, USA$^\ast$}
\affiliation{The University of Texas at Austin, Austin, TX 78712, USA$^\ast$}
\affiliation{E\"otv\"os Lor\'and University, Budapest, 1117 Hungary$^\ast$}
\affiliation{University of Adelaide, Adelaide, SA 5005, Australia$^\ast$}
\affiliation{University of Szeged, 6720 Szeged, D\'om t\'er 9, Hungary$^\ast$}
\affiliation{Embry-Riddle Aeronautical University, Prescott, AZ   86301 USA$^\ast$}
\affiliation{National Institute for Mathematical Sciences, Daejeon 305-390, Korea$^\ast$}
\affiliation{Perimeter Institute for Theoretical Physics, Ontario, Canada, N2L 2Y5$^\ast$}
\affiliation{National Astronomical Observatory of Japan, Tokyo  181-8588, Japan$^\ast$}
\affiliation{Universitat de les Illes Balears, E-07122 Palma de Mallorca, Spain$^\ast$}
\affiliation{Korea Institute of Science and Technology Information, Daejeon 305-806, Korea$^\ast$}
\affiliation{University of Southampton, Southampton, SO17 1BJ, United Kingdom$^\ast$}
\affiliation{Institute of Applied Physics, Nizhny Novgorod, 603950, Russia$^\ast$}
\affiliation{Lund Observatory, Box 43, SE-221 00, Lund, Sweden$^\ast$}
\affiliation{Hanyang University, Seoul 133-791, Korea$^\ast$}
\affiliation{Seoul National University, Seoul 151-742, Korea$^\ast$}
\affiliation{University of Strathclyde, Glasgow, G1 1XQ, United Kingdom$^\ast$}
\affiliation{Southern University and A\&M College, Baton Rouge, LA  70813, USA$^\ast$}
\affiliation{University of Rochester, Rochester, NY  14627, USA$^\ast$}
\affiliation{Rochester Institute of Technology, Rochester, NY  14623, USA$^\ast$}
\affiliation{Hobart and William Smith Colleges, Geneva, NY  14456, USA$^\ast$}
\affiliation{University of Sannio at Benevento, I-82100 Benevento, Italy and INFN (Sezione di Napoli), Italy$^\ast$}
\affiliation{Louisiana Tech University, Ruston, LA  71272, USA$^\ast$}
\affiliation{McNeese State University, Lake Charles, LA 70609 USA$^\ast$}
\affiliation{University of Washington, Seattle, WA, 98195-4290, USA$^\ast$}
\affiliation{Andrews University, Berrien Springs, MI 49104 USA$^\ast$}
\affiliation{Trinity University, San Antonio, TX  78212, USA$^\ast$}
\affiliation{Southeastern Louisiana University, Hammond, LA  70402, USA$^\ast$}
\author{J.~Abadie$^\text{1}$}\noaffiliation\author{B.~P.~Abbott$^\text{1}$}\noaffiliation\author{R.~Abbott$^\text{1}$}\noaffiliation\author{T.~D.~Abbott$^\text{2}$}\noaffiliation\author{M.~Abernathy$^\text{3}$}\noaffiliation\author{T.~Accadia$^\text{4}$}\noaffiliation\author{F.~Acernese$^\text{5a,5c}$}\noaffiliation\author{C.~Adams$^\text{6}$}\noaffiliation\author{R.~Adhikari$^\text{1}$}\noaffiliation\author{C.~Affeldt$^\text{7,8}$}\noaffiliation
% ...

\collaboration{$^\ast$The LIGO Scientific Collaboration and $^\dagger$The Virgo Collaboration}
\noaffiliation




%
%
\date[\relax]{ RCS \thercsid; compiled \today }
\fake{}

\begin{abstract}
We report on a search for gravitational waves from coalescing compact binaries
using LIGO and Virgo observations between July 7, 2009 and October 20, 2010.
We searched for signals from binaries with total mass between $2$ and
$25~\Msun$; this includes binary neutron stars, binary black holes, and
binaries consisting of a black hole and neutron star.  The detectors were
sensitive to systems up to \BNShd\,Mpc distant for binary neutron stars, and
further for higher mass systems. No gravitational-wave signals were detected.
We report upper limits on the rate of compact binary coalescence as a function
of total mass, including the results from previous LIGO and Virgo observations.
The cumulative 90\%-confidence rate upper limits of the binary coalescence of
binary neutron star, neutron star--black hole and binary black hole systems
are \BNSul, \NSBHul\ and \BBHul\,$\mathrm{Mpc^{-3}yr^{-1}}$, respectively.
These upper limits are up to a factor 1.4 lower than previously derived limits.
We also report on results from a blind injection challenge.
\end{abstract}

\maketitle


\section{Introduction}\label{sec:overview}

\acrodef{BBH}{binary black holes}
\acrodef{BNS}{binary neutron stars}
\acrodef{NSBH}{neutron star--black hole binaries}
\acrodef{SNR}{signal-to-noise ratio}
\acrodef{LIGO}{Laser Interferometer Gravitational-wave Observatory}
\acrodef{LHO}{LIGO Hanford Observatory}
\acrodef{LLO}{LIGO Livingston Observatory}
\acrodef{LSC}{LIGO Scientific Collaboration}
\acrodef{CBC}{compact binary coalescence}
\acrodef{FAR}{false alarm rate}
\acrodef{VSR1}{the first Virgo science run}
\acrodef{VSR2}{Virgo's second science run}
\acrodef{VSR3}{its third science run}
\acrodef{S5}{LIGO's fifth science run}
\acrodef{S6}{LIGO's sixth science run}

During 2009 and 2010, both the \ac{LIGO}~\cite{Abbott:2007kv} and
Virgo~\cite{Acernese:2008b} gravitational-wave detectors undertook science runs
with better sensitivity across a broader range of frequencies than previously
achieved.  Among the most promising sources of gravitational waves for these
detectors are compact stellar mass binaries as they spiral in toward each other
and merge.  For such systems, which include \ac{BNS}, \ac{BBH}, and \ac{NSBH},
the late stages of inspiral and merger occur in the most sensitive band
(between 40 and 1000\,Hz) of the \ac{LIGO} and Virgo detectors.  In this paper,
we report on a search for gravitational waves from binary systems with a
maximum total mass of $25\,\Msun$, and a minimum component mass of $1\,\Msun$.  

A hardware injection was performed during the data collection without the
knowledge of the data analysis teams as part of a ``blind injection
challenge"~\cite{GW100916web}. This challenge was intended to test the data
analysis procedures and processes for evaluating candidate events. The
injection was performed by coherently actuating the mirrors on the \ac{LIGO}
and Virgo detectors to mimic a gravitational-wave signal. Prior to its
unveiling as an injection (``unblinding"), the event was determined to be a
candidate gravitational wave: it was found to have a false alarm rate of less
than \dogFAR~and no evidence for an instrumental or environmental origin could
be found.  After the analysis of the event was finished it was revealed to be a
blind injection and removed from the data.

With the blind injection removed there were no gravitational waves observed
above the noise background.  As a result we place upper limits on rates of
\ac{CBC}, using upper limits from previous gravitational-wave searches
\cite{\sfivelvc} as prior information. The upper limits presented here are up
to a factor 1.4 lower than previously derived limits but still two to three
orders of magnitude above expected \ac{CBC} rates \cite{ratesdoc}.

The paper is laid out as follows.  In Section \ref{sec:dets}, we provide
a brief description of the detectors and their sensitivities during
\ac{S6} and Virgo's second and third science runs. In
Section \ref{sec:search} we present a brief overview of the analysis
methods used in performing the search.  In Section~\ref{sec:blindinjection}
we discuss the recovery of the blind injection.
In Section \ref{sec:results} we
present the results of the search with the blind injection removed.
In Section
\ref{sec:ul} we give the upper limits obtained from the search and close
with a brief discussion in Section \ref{sec:discussion}.

\section{Detectors}
\label{sec:dets}

The \ac{LIGO} observatory comprises two sites, one in Hanford, WA and the
second in Livingston, LA.  The data used in this search were taken during
\ac{S6}, which took place between 7 July 2009 and 20 October 2010.  During
\ac{S6} each of these sites operated a single 4km laser interferometer, denoted
as H1 and L1 respectively.  The 2km H2 instrument at the Hanford site which
operated in earlier science runs was not operational in \ac{S6}.  Following
\ac{S5} \cite{Abbott:2007kv}, several hardware changes were made to the
\ac{LIGO} detectors so that prototypes of advanced LIGO technology could be
installed and tested~\cite{Adhikari:2006,Smith2009}. 
This included the installation of a higher power
laser, and the implementation of a DC readout system that
included a new output mode cleaner on an advanced LIGO seismic isolation
table~\cite{DCreadoutELIGO}. 
In addition, the hydraulic seismic isolation system
was improved by fine tuning its feed-forward path.  

The Virgo detector (denoted V1) is a single, 3km laser interferometer located
in Cascina, Italy.  The data used in this search were taken from both
\ac{VSR2},  which ran from 7 July 2009 to 8 January 2010, and \ac{VSR3},  which
ran from 11 August 2010 to 20 October 2010.  In the period between \ac{VSR1}
and \ac{VSR2}, several enhancements were made to the Virgo detector.
Specifically, a more powerful laser was installed in Virgo, along with a
thermal compensation system and improved scattered light mitigation.  During
early 2010, monolithic suspensions were installed, which involved replacing
Virgo's test masses with new mirrors hung from fused-silica fibers
\cite{Virgo:2010cqg}.  VSR3 followed this upgrade.

\begin{figure}[tp]
\vskip -0.4cm
\includegraphics[width=1.06\columnwidth]{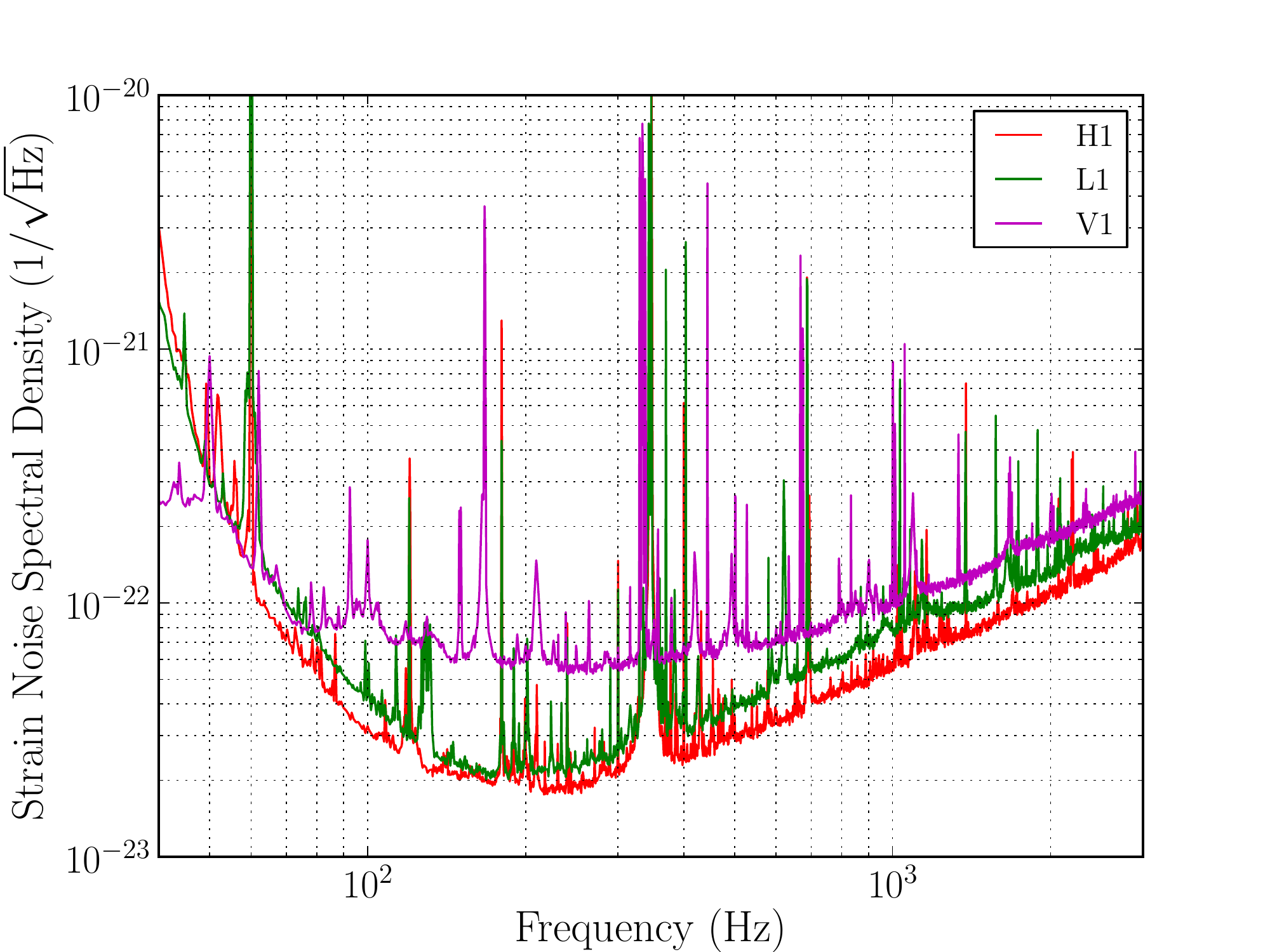}
\caption{Typical detector strain noise spectral density for the LIGO S6
and Virgo VSR2/3 runs. From lowest to highest at $10^2$ Hz, the curves are for
the H1, L1 and V1 detectors.
\label{fig:noise_curve}}
\end{figure}

\begin{figure}[t]
\hspace*{-2.5mm}
\includegraphics[width=0.97\columnwidth]{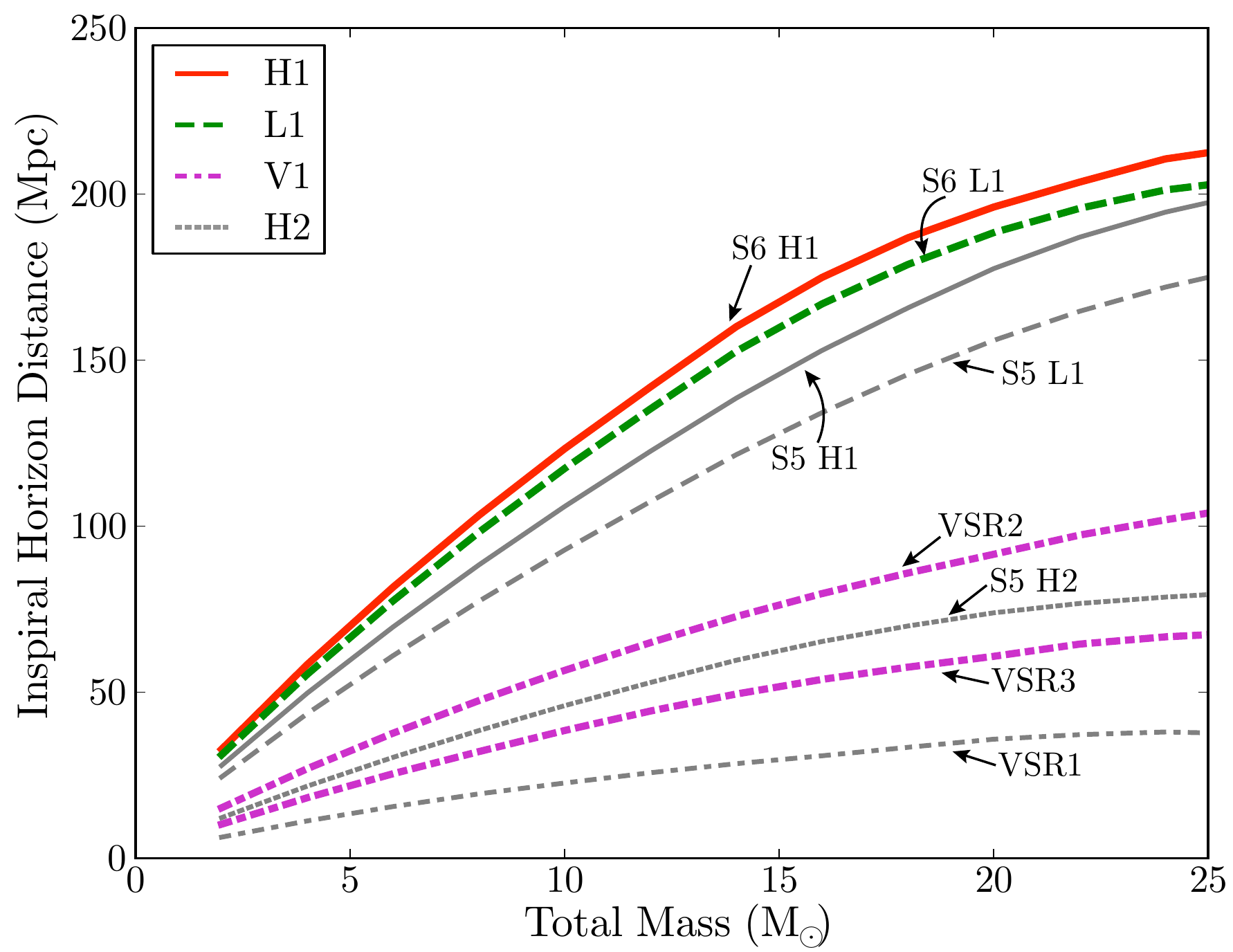}
\caption{
Inspiral horizon distance versus the total mass of equal-mass binaries from
S5/VSR1 (gray lines) and S6/VSR2/VSR3 (colored lines). The horizon distance is
the distance at which an optimally located and oriented binary would produce an
expected signal-to-noise ratio of 8.  The figure shows the best sensitivity
achieved by each detector during the runs.} 
\label{fig:sensitivity}
\end{figure}

The sensitivity of the detectors during the \ac{S6}, \ac{VSR2} and
\ac{VSR3} runs is shown in Figure \ref{fig:noise_curve}.  The
corresponding sensitivity to binary coalescence signals is shown in
Figure \ref{fig:sensitivity}.  This figure shows the distance at which
an optimally oriented and located binary would produce a \ac{SNR} of 8
in a given detector.  The figure illustrates the improvement in
sensitivity for the \ac{LIGO} detectors between \ac{S5} and \ac{S6} and
for Virgo between \ac{VSR1} and \ac{VSR2}.  The reduction in the horizon
distance of the Virgo detector in \ac{VSR3} is due to a mirror with an
incorrect radius of curvature being installed during the conversion to
monolithic suspension.

\section{Binary Coalescence Search}
\label{sec:search}

To search for gravitational waves from compact binary
coalescence~\cite{Collaboration:2009tt,Abbott:2009qj, S5LowMassLV}, we use
matched filtering to correlate the detector's strain output with a theoretical
model of the gravitational waveform~\cite{Allen:2005fk}. Each detector's output
is separately correlated against a bank~\cite{BBCCS:2006} of template waveforms
generated at 3.5 post-Newtonian order in the frequency
domain~\cite{Blanchet:1995ez, Blanchet:2004ek}. Templates were laid out across
the mass range such that no more than $3\%$ of the \ac{SNR} was lost due to the
discreteness of the bank. Only non-spinning waveforms with zero eccentricity
and a component mass $\geq 1\,\Msun$ were generated, and the templates were
terminated prior to merger. In the early stages of the run, as in previous
searches \cite{\sfive1yr,\12to18,\sfivelvc}, the template bank included
waveforms from binaries with a total mass $M \leq 35\,\Msun$. However, the
search results indicated that the higher mass templates ($M > 25\,\Msun$) were
more susceptible to non-stationary noise in the data. Furthermore, it is at
these higher masses where the merger and ringdown phases of the signal come
into the detectors' sensitive bands. Consequently, the upper mass limit of this
search was reduced to $25\,\Msun$ during the latter stages of the science run.
Results of a search for higher mass binary black holes using non-spinning,
full coalescence (inspiral-merger-ringdown) template waveforms, such as
in~\cite{Collaboration:S5HighMass}, will be presented in a future publication.
Although the template waveforms in this search neglect the spin of the binary
components, the search is still capable of detecting binaries whose waveforms
are modulated by the effect of spin~\cite{VanDenBroeck:2009gd}. 

We require candidate signals to have a matched filter \ac{SNR} greater than 5.5 
in at least two detectors, and to have consistent values of template masses and 
time of arrival (allowing for travel-time difference) across the detectors 
where this threshold is exceeded~\cite{Robinson:2008}. We use a chi-squared 
test~\cite{Allen:2004} to suppress non-Gaussian noise transients, which have a 
high \ac{SNR} but whose time-frequency evolution is inconsistent with the 
template waveform. If the reduced chi-squared of a signal, $\chi^2_r$, is greater 
than unity, we re-weight the \ac{SNR} $\rho$ in order to suppress the 
significance of false signals, obtaining a re-weighted \ac{SNR} statistic\footnote{
Equation \ref{eq:new_snr} is an improvement over the ``effective \ac{SNR}" used
to rank events in \cite{\sfive1yr,\12to18,\sfivelvc}. Most notably: while effective
\ac{SNR} also re-weighted \ac{SNR} using $\chi^2_r$, it became larger than \ac{SNR}
when $\chi_r^2 < 1$. This made it susceptible to over weighting events
that had statistical downward fluctuations in $\chi_r^2$.
}
\begin{equation}\label{eq:new_snr} \hat{\rho} = \begin{cases} {\displaystyle
\frac{\rho}{[(1+(\chi^2_r)^3)/2]^{1/6}}} & \mbox{for } \chi^2_r > 1, \\ \rho &
\mbox{for } \chi^2_r \leq 1.  \end{cases}  \end{equation}
Our analysis reports the coalescence time and the quadrature sum, $\rho_c$, of
re-weighted SNRs for events coincident between the detectors. The statistic
$\rho_c$ is then used to rank events by their significance above the expected
background.  To measure the background rate of coincident events in the search,
we time-shift data from the detectors by an amount greater than the
gravitational-wave travel time difference between detector sites and re-analyze
the data. Many independent time-shifts are performed to obtain a good estimate
of the probability of accidental coincidence of noise transients at two or more
sites.  The analysis procedure described above is similar to the one used in
previous searches of \ac{LIGO} and Virgo data, such as \cite{\sfive1yr} and
\cite{\12to18}; it will be described in more detail in \cite{ihopePaper:2012}.

The background rates of coincident events were initially estimated using
100 time-shifted analyses. These background rates vary depending on the
binary's mass --- via the waveform duration and frequency band --- and also on
the detectors involved in the coincidence (the {\it event type}).
The relevant mass parameter is the binary's chirp mass, $\mathcal{M} \equiv 
(m_1 m_2)^{3/5} (m_1+m_2)^{-1/5}$,
where $m_1$ and $m_2$ are the component masses in the binary system.
Thus, we sort coincident events into three bins by chirp mass, 
and by event type \cite{Collaboration:2009tt}. 

The requirement of a coincident signal between at least two sites 
restricts the times that can be analyzed to four distinct types of 
\emph{coincident time}. Between July 2009 and October 
2010, a total of \totalpreVeto~of two-or-more-site coincident data was 
collected.  This comprised \HLVltpreVeto~of H1L1V1 coincident data, 
\HLltpreVeto~of H1L1 data, \HVltpreVeto~of H1V1 data, and 
\LVltpreVeto~of L1V1 data. During H1L1V1 coincident time 
there are four distinct event types: H1L1V1, 
H1L1, H1V1, and L1V1. In S6/VSR2, all four event types were kept. 
In S6/VSR3, H1V1 and L1V1 events in triple-coincident time were 
discarded due to the heightened rate of transient noise artifacts in Virgo 
and its decreased sensitivity. 

For each candidate, a \ac{FAR} is computed by comparing its $\rho_c$ 
value to background events in the same mass bin and coincident time 
and with the same event type. 
Candidates' \ac{FAR} values are then compared to background events in 
{\em all}\/ bins and event types, over the appropriate coincident time, 
to calculate a combined \ac{FAR}. This is the detection statistic which is 
used to assess the significance of events over the entire analysis time.

Due to the finite number of time-shifts performed, the
smallest non-zero FAR that can be calculated is $1/T_{\rm bg}$, where
$T_{\rm bg}$ is the total background time obtained by summing the
coincident live time in each time-shift. If an event was found to be
louder than all background events within its analysis period, additional 
time-shifted analyses were performed to calculate a more precise \ac{FAR} 
for the event. 

Although the detectors are enclosed in vacuum systems and isolated from
vibrational, acoustic and electromagnetic disturbances, their typical output
data contains a larger number of transient noise events (glitches) with higher
amplitude than expected from Gaussian processes alone.  Each observatory is
equipped with a system of environmental and instrumental monitors that are
sensitive to glitch sources but have a negligible sensitivity to gravitational
waves. These sensors were used to identify times when the detector output was
potentially corrupted~\cite{Christensen:2010,Robinet:2010zz,SeisVeto,VirgoDetChar}.  We
grouped these times into two categories: periods with well-understood couplings
between non-gravitational-wave sources and detector output, and periods when a
statistical correlation was found but a coupling mechanism was not identified.
In our primary search --- which included the identification of
gravitational-wave candidates and the calculation of upper limits --- we
removed (\emph{vetoed}) times that fell in either of the two categories from
the analysis, along with any coincident events that occurred during these
periods.  We also performed a secondary search for possible loud candidate
events, in which only the times with known couplings were vetoed.

Approximately $10\%$ of the data, designated {\it playground}, was used for
tuning and data quality investigations.  This data was searched for
gravitational waves, but not used in calculating upper limits.  After all
vetoes were applied and playground time excluded, there was \HLVlt~of H1L1V1
time, \HLlt~of H1L1 time, \HVlt~of H1V1 time, and \LVlt~of L1V1 time, giving a
total analysis time of \totalTime.

A substantial change from the analysis procedure of \cite{\12to18} was that
data were analyzed in two-week blocks with a latency of two to four weeks, to
allow for feedback of information to ongoing detector characterization efforts
and to improve data quality. Thus, during the search many new vetoes were
introduced resulting from improved understanding of the detectors. However,
significant numbers of delta-function-like glitches with large amplitudes
remained unvetoed in the LIGO detectors. These were found to cause artifacts in
the matched filter output over a short time surrounding the glitch: thus,
during the latter stages of the search, 8\,s of time on either side of any
matched filter SNR exceeding 250 was vetoed. Times removed from the primary
search by this veto were still examined for possible loud events.

\section{Blind injection recovery}
\label{sec:blindinjection}

\begin{figure}[tp]
\vskip 0.1cm
\hspace*{-1.5mm}
\includegraphics[width=\columnwidth]{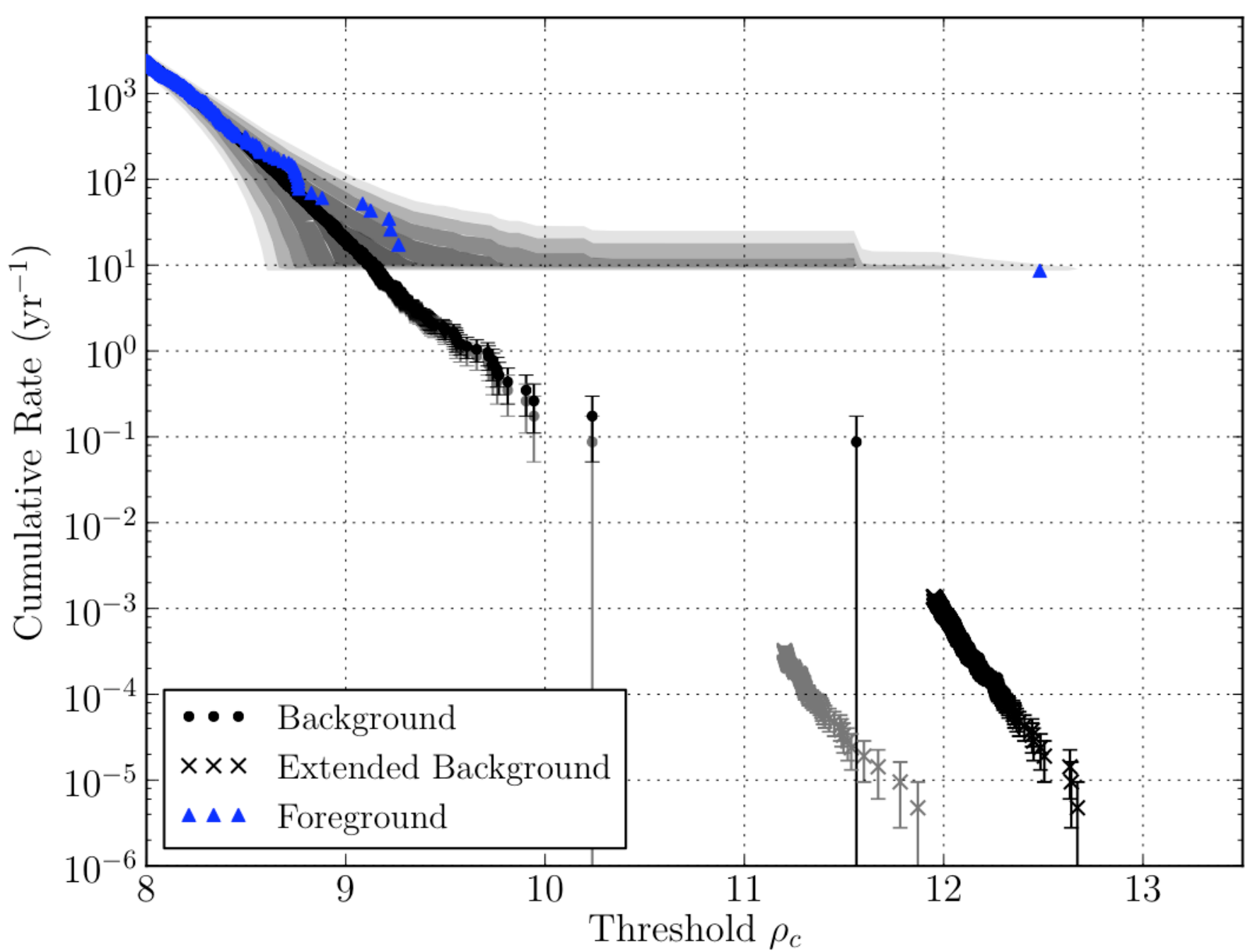}
\caption{The cumulative rate of events with chirp mass $3.48 \le
\mathcal{M}/\Msun < 7.40$ coincident in the H1 and L1 detectors, seen in
four months of data around the 16 September candidate, as a function of
the threshold ranking statistic $\rho_c$.  The blue triangles show
coincident events.  Black dots show the background estimated from 100
time-shifts. Black crosses show the extended background estimation from
all possible 5-second shifts on this data restricted, for computational
reasons, to only the tail of loudest events.  The gray dots and crosses
show the corresponding background estimates when 8 seconds of data
around the time of the candidate are excluded.  Gray shaded contours
show the $1 - 5\sigma$ (dark to light) consistency of coincident events
with the estimated background including the extended background
estimate, for the events and analysis time shown, including the
candidate time.  This event was later revealed to have been a blind
injection. }
\label{fig:far}
\end{figure}

The search pipeline described above identified a gravitational-wave
candidate occurring on \dogDate~at \recoveredDogTime, with \dogSNR\ in
coincidence between the two LIGO detectors in the middle mass bin $3.48 
\le \mathcal{M}/\Msun < 7.40$.  The highest matched-filter SNR
obtained in the search was $15$ at $\mathcal{M} = 4.7\,\Msun$ in H1 and
$10$ at $\mathcal{M} = 4.4\,\Msun$ in L1.  This difference in SNRs is
consistent with typical differences in antenna response factors for
these differently-oriented detectors.  Virgo was also operating at the
time of the event, but its sensitivity was a factor of approximately
four lower than the LIGO detectors; the absence of a signal in Virgo
above the single-detector SNR threshold of 5.5 was consistent with this
fact.  In the LIGO detectors, the signal was louder than all
time-shifted H1L1 coincident events in the same mass bin throughout
\ac{S6}. However, with only 100 time-shifts, we could only bound the FAR
to $< 1/23$ years, even when folding in all data from the entire
analysis. To obtain a better estimate of the event's FAR we performed
all possible multiples of 5-second time-shifts on four calendar months
of data around the event, corresponding to an effective analysis time of
$2.0\times 10^5$ years.  We found five events with a value of $\rho_c$
equal to or larger than the candidate's, as shown in
Figure~\ref{fig:far}.  These five events were all coincidences between
the candidate's signal in H1 and time-shifted transient noise in L1.
When we excluded 8 seconds from around the event's time in the
background estimation, we found {\em no}\/ background events with
$\rho_c$ greater than the candidate and we obtained a significantly
different background distribution, also shown in Figure~\ref{fig:far}. 

Including the events at the time of the candidate in the background
estimate, the FAR of the event in the $3.48 \le \mathcal{M}/\Msun <
7.40$ mass bin, coincident in the LIGO detectors, was estimated to be
\dogTrialFAR.  Since this event occurred in H1L1V1 time during VSR3,
only two event types were considered: H1L1 double-coincident
events and H1L1V1 triple-coincident events.  This resulted in a trials
factor of 6 (accounting for the three mass bins and two coincidence
types) and a combined FAR of \dogFAR. The false alarm probability of 
this event in this analysis, over the 0.47\,yr of coincident time 
remaining after all vetoes were applied, was \dogFAP.

The detectors' environmental monitoring channels record data from seismometers,
accelerometers, microphones, magnetometers, radio receivers, weather sensors,
and a cosmic ray detector. Injections of environmental signals and other tests
indicate that these channels are much more sensitive to environmental signals
than the gravitational wave readout channels are. Arrays of these detectors
were operating and providing full coverage at the time of the event, and did
not record environmental signals that could account for the event.
Environmental signal levels at our observatories and at external
electromagnetic “weather” observatories were typical of quiet times. Mechanisms
that could cause coincident signals among widely separated detectors --- such
as earthquakes, microseismic noise due to large weather systems, and
electromagnetic disturbances in the ionosphere~\cite{Christensen:1992,
Singh:2004} ---  were therefore ruled out. 

A loud transient occurred in L1 9 seconds before the coalescence time of the
signal. That transient belonged to a known family of sharp ($\sim$\,10\,ms) and
loud (SNR $\approx 200$-$80000$) glitches that appear 10--30 times per day in
the output optical sensing system of this detector. Since the candidate signal
swept through the sensitive band of the detector, from 40\,Hz to coalescence,
in less than 4 seconds, it did not overlap the loud transient.  Studies,
including re-analysis of the data with the glitch removed, indicated that the
signal was not related to the earlier instrumental glitch.  No evidence was
found that the observed signal was associated with, or corrupted by, any
instrumental effect.  

Following the completion of this analysis, the event was revealed to be a blind
injection. While the analysis groups did not know the event was an injection
prior to its unblinding, they did know that one or more blind injections may be
performed during the analysis period. Such blind injections have been carried
out before: see~\cite{S5LowMassLV} for the results of a blind injection
performed in a previous run. This event was the only coherent \ac{CBC} blind
injection performed during S6 and VSR2 and 3. The injection was
identified as a gravitational-wave candidate with high probability, and the blind
injection challenge was considered to be successful \cite{GW100916web}. 

In order to more accurately determine the parameters of the event prior to the
unblinding, we performed coherent Bayesian analyses of the data using models of
both spinning and non-spinning compact binary objects
\cite{Veitch:2010,Feroz:2009,Sluys:2008a,Sluys:2008b,Roever2007}.  These
analyses showed evidence for the presence of a weak signal in Virgo, consistent
with the signal seen by the two LIGO detectors.  The strength of a signal in
Virgo is an important input to the localization of a source in the sky.
Parameter estimates varied significantly depending on the exact model used for
the gravitational waveform, particularly when we included spin effects.
However, conservative unions of the confidence intervals from the different
waveform models were consistent with most injected parameters, including chirp
mass, time of coalescence, and sky location. In addition, the signal was
correctly identified as having at least one highly-spinning component with the
spin misaligned with the angular orbital momentum. We will describe the details
of parameter estimation on this and other \ac{CBC} injections in a future paper
(in preparation).

\section{Search results}
\label{sec:results}

After the event was revealed to be a blind injection the data containing it was
removed from the analysis. With the injection excluded, there were no
gravitational-wave candidates observed in the data. Indeed the search result
was consistent with the background estimated from time-shifting the data.  The
most significant event was an L1V1 coincidence in L1V1 time with a
combined FAR of \firstFAR. The second and third most significant events had
combined FARs of \secondFAR~and \thirdFAR, respectively. All of these events
were consistent with background: having analyzed
$\mathrm{\sim0.5\,yr}$~of data, we would expect the loudest event to
have a FAR of \expectedLoudestFAR. Although no detection candidates were found,
a detailed investigation of the loudest events in each analysis period was
performed, to improve our understanding of instrumental data quality.

\section{Binary Coalescence Rate Limits}
\label{sec:ul}

Given the absence of gravitational-wave signals, we used our observations
to set upper limits on coalescence rates of \ac{BNS}, \ac{BBH}, and
\ac{NSBH} systems.  We used the procedure described in
\cite{Fairhurst:2007qj,loudestGWDAW03,Biswas:2007ni} to compute Bayesian 90\%\ 
confidence level upper limits on the coalescence rate for the various systems, making 
use of previous results \cite{\sfive1yr,\12to18,\sfivelvc} as prior information on the
rates.

The rate of binary coalescences in a spiral galaxy is expected to be
proportional to the star formation rate, and hence blue light
luminosity, of the galaxy \cite{LIGOS3S4Galaxies}.  Previous searches
\cite{\sfive1yr,\12to18,\sfivelvc} presented upper limits in terms of
blue light luminosity, using units of $\mathrm{L_{10}^{-1}yr^{-1}}$,
where one $\mathrm{L_{10}}$ is $10^{10}$ times the solar blue light
luminosity.  There are, however, numerous challenges to evaluating the
upper limit as a function of luminosity, not least due to the large
uncertainties in both the luminosity of and distance to nearby galaxies,
as well as the lack of a complete galaxy catalogue at larger distances
\cite{Fairhurst:2007qj, LIGOS3S4Galaxies}.  On large scales (greater
than $\sim 20\,\mathrm{Mpc}$), the luminosity per unit volume is
approximately constant; consequently the analysis can be simplified by
reporting upper limits per unit volume per unit time.  During the
current analysis, the sensitivity of the detectors to the systems of
interest (as shown in Figure \ref{fig:sensitivity}) was sufficiently
large that we could assume signals were uniformly distributed in volume.
We therefore quote upper limits in units of $\mathrm{Mpc^{-3}yr^{-1}}$.
To incorporate the previous results as prior distributions, we converted
from $\mathrm{L_{10}}$ to $\mathrm{Mpc^3}$ using a conversion factor of
$0.02 \, \mathrm{L_{10}}$ per $\mathrm{Mpc^3}$ \cite{LIGOS3S4Galaxies}.

We estimate the volume to which the search is sensitive by reanalyzing
the data with the addition of a large number of simulated signals 
(``software injections'') in order to model the source population. 
Our ability to detect a signal depends
upon the parameters of the source, including the component masses, the
distance to the binary, its sky location, and its orientation with
respect to the detectors. Numerous signals with randomly chosen
parameters were therefore injected into the data. To compute the
sensitive volume for a given binary mass, we perform a Monte Carlo
integration over the other parameters to obtain the efficiency of the
search---determined by the fraction of simulated signals found louder
than the loudest foreground event---as a function of distance.
Integrating the efficiency as a function of distance then gives the 
sensitive volume.

We consider several systematic uncertainties that limit the accuracy of
the measured search volume and therefore the upper limits
\cite{\sfive1yr}: detector calibration errors (conservatively estimated to
be 14\%\ in sensitive distance combined over all three detectors and over the entire
observational period, and a 2\%\ bias correction), 
waveform errors (taken to be a one-sided 10\% \cite{Fairhurst:2007qj} 
bias towards lower sensitive distance),
and Monte Carlo statistical errors (3-5\%\ in sensitive volume).  
We convert the sensitive distance uncertainties to volume uncertainties,
and then marginalize over the uncertainty in volume 
to obtain an upper limit which takes into account these
systematic uncertainties \cite{Fairhurst:2007qj}.

\begin{table}
\center
\begin{tabular}{p{3.7cm} | c  c  c}
\hline \hline
\raggedright{System}                                                    &   BNS       &    NSBH    &    BBH      \\[2pt]
\hline                                                                                             
\raggedright{Component masses ($\Msun$)}                                & 1.35 / 1.35 & 1.35 / 5.0 & 5.0 / 5.0   \\[2pt]
\raggedright{$D_{\rm horizon}$ ($\mathrm{Mpc}$)}                        &   \BNShd    &   \NSBHhd  &   \BBHhd    \\[2pt]
\raggedright{Non-spinning upper limit ($\mathrm{Mpc^{-3} yr^{-1}}$)}    &   \BNSul    &   \NSBHul  &   \BBHul    \\[2pt]
\raggedright{Spinning upper limit ($\mathrm{Mpc^{-3} yr^{-1}}$)}        &    $\cdots$     &   \sNSBHul &   \sBBHul   \\[2pt]
\hline
\hline
\end{tabular}
\caption{Rate upper limits of \ac{BNS}, \ac{BBH} and \ac{NSBH} coalescence,
assuming canonical mass distributions.  $D_{\rm horizon}$ is the horizon
distance averaged over the time of the search.  
The sensitive distance averaged over all sky locations and binary orientations
is $D_{\rm avg}\simeq D_{\rm horizon}/2.26$~\cite{FinnChernoff:1993}.
The first set of upper limits
are those obtained for binaries with non-spinning components.  The second set
of upper limits are produced using black holes with a spin uniformly
distributed between zero and the maximal value of $Gm^{2}/c$.}
\label{tab:ul}
\end{table}

\begin{figure}[htb]
\begin{center}
\vskip -0.5cm
\hspace*{-2mm}
\includegraphics[width=1.04\columnwidth]{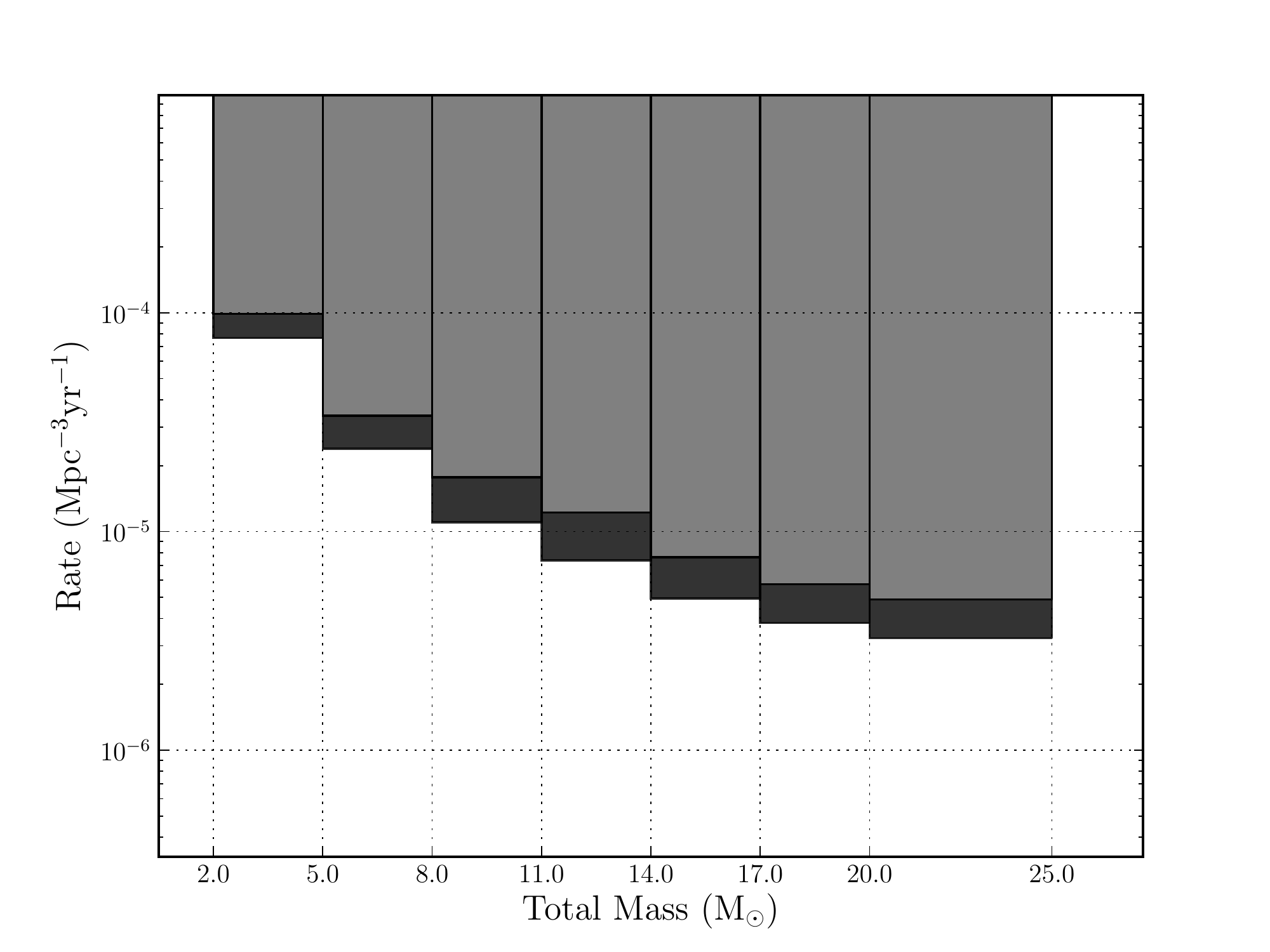} \\
\vskip -0.2cm 
\hspace*{-2mm}
\includegraphics[width=1.04\columnwidth]{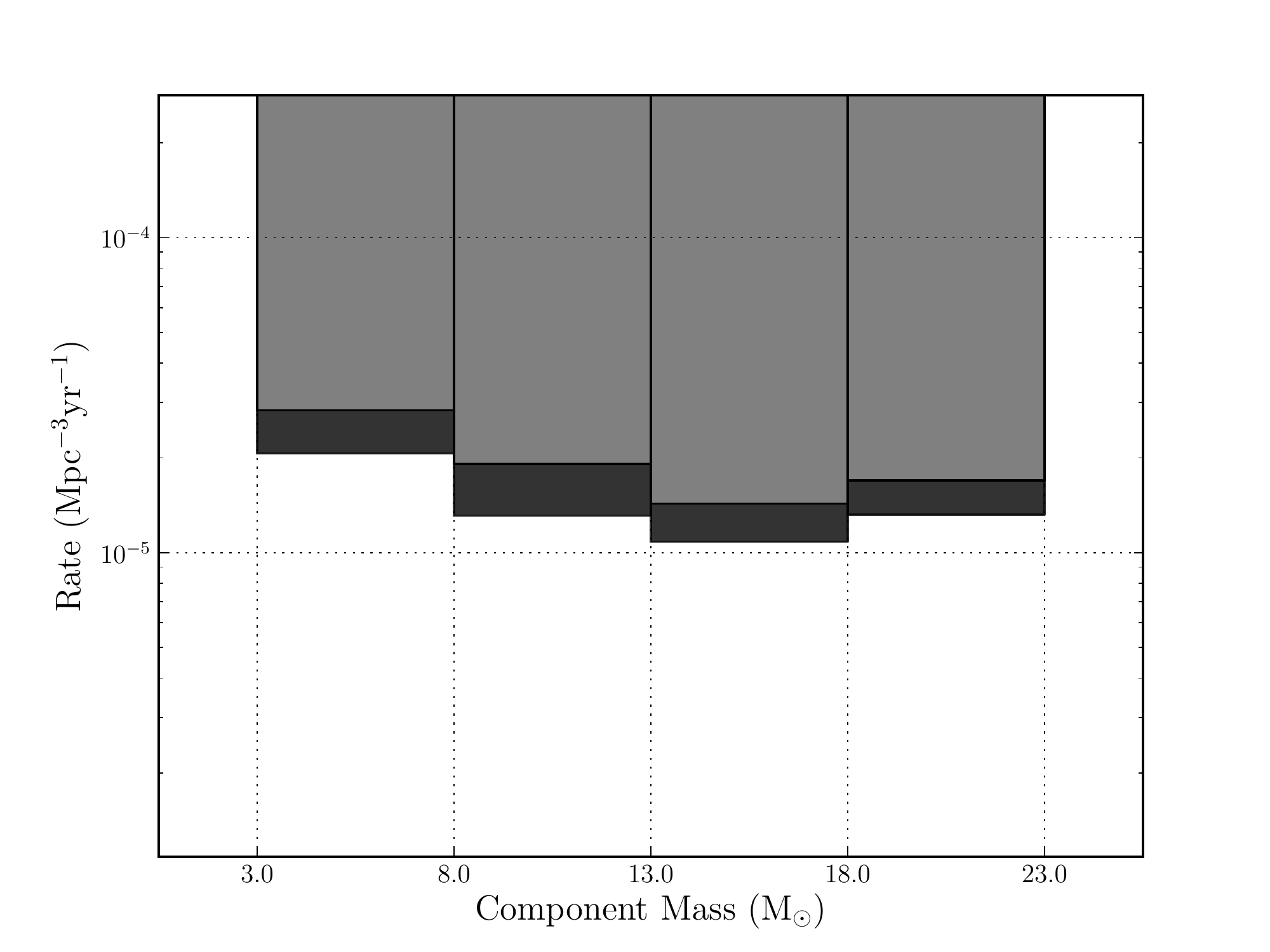}
\caption{The marginalized upper limits as a function of mass. The top plot
shows the limit as a function of total mass $M$, using a distribution uniform in 
$m_1$ for a given $M$. The lower plot shows the limit as a function of the black hole
mass, with the neutron star mass restricted to the range $1-3\,\Msun$. The
light bars indicate upper limits from previous searches. The dark bars
indicate the combined upper limits including the results of this search.}
\label{fig:ULplots}
\end{center}
\end{figure}

\begin{figure}[ht]
\vskip -0.5cm
\hspace*{-2mm}
\includegraphics[width=1.04\columnwidth]{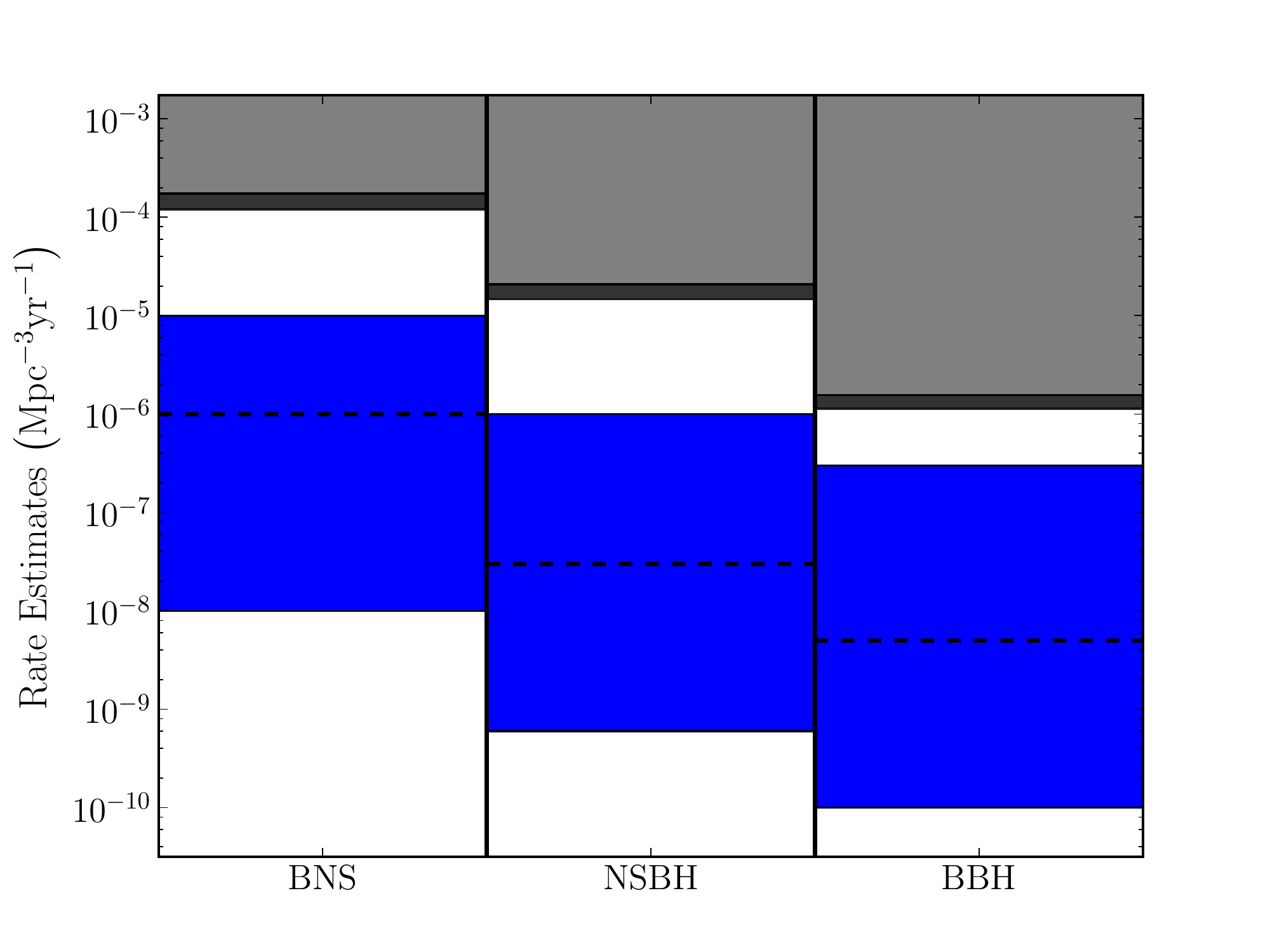}
\caption{Comparison of CBC upper limit rates for \ac{BNS}, \ac{NSBH} and
\ac{BBH} systems.  The light gray regions display the upper limits obtained in
the S5-VSR1 analysis; dark gray regions show the upper limits obtained in this
analysis, using the S5-VSR1 limits as priors.  The new limits are up to a
factor of 1.4 improvement over the previous results.  The lower (blue) regions
show the spread in the astrophysically predicted rates, with the dashed-black
lines showing the ``realistic'' estimates \cite{ratesdoc}. {\it Note}: In
\cite{ratesdoc}, \ac{NSBH} and \ac{BBH} rates were quoted using a black-hole
mass of $10\,\Msun$. We have therefore rescaled the \ac{S5} and \ac{S6}
\ac{NSBH} and \ac{BBH} upper limits in this plot by a factor of
$(\mathcal{M}_{5} / \mathcal{M}_{10})^{5/2}$, where $\mathcal{M}_{10}$ is the
chirp mass of a binary in which the black hole mass is $10\,\Msun$ and
$\mathcal{M}_5$ is the chirp mass of a binary in which the black hole mass is
$5\,\Msun$.}
\label{fig:rate_comp}
\end{figure}

In Table \ref{tab:ul} we present the marginalized upper limits at the
$90\%$ confidence level assuming canonical mass distributions for
non-spinning \ac{BNS} ($m_1 = m_2 = 1.35 \pm 0.04\,\Msun$), \ac{BBH}
($m_1 = m_2 = 5 \pm 1\,\Msun$), and \ac{NSBH} ($m_1 = 1.35 \pm 0.04\,\Msun$, 
$m_2 = 5 \pm 1\,\Msun$) systems. We also compute upper limits as a
function of total mass $M$, using an injection population distributed 
uniformly over $M$ and uniformly over $m_1$ for a given $M$. For
\ac{NSBH} systems we present the upper limit as a function of black hole
mass, keeping the neutron-star mass fixed in the range $1-3\,\Msun$. These
are presented in Figure \ref{fig:ULplots}. Figure \ref{fig:rate_comp}
compares the upper limits obtained in this analysis (dark gray
regions) to limits obtained in our previous searches up to
S5/VSR1~\cite{\sfivelvc} (light gray region) and to
astrophysically-predicted rates (blue regions) for \ac{BNS}, \ac{NSBH},
and \ac{BBH} systems. The improvement over the previous limits is up to 
a factor of 1.4,
depending on binary mass; this reflects the additional observation time
and improved sensitivity of the S6/VSR2/VSR3 data with respect to 
all previous observations.

Although we searched with a bank of non-spinning templates, we compute
upper limits for \ac{NSBH} and \ac{BBH} systems in which one or both of
the component masses are spinning. These results are also presented in
Table \ref{tab:ul}. We did not compute upper limits for spinning
\ac{BNS} systems because astrophysical observations indicate that
neutron stars cannot have large enough spin to significantly affect
waveforms observable in the \ac{LIGO} frequency band
\cite{ATNF:psrcat,Apostolatos:1994}. Black hole spins were uniformly
distributed in both orientation and magnitude, $S$, with $S$ constrained
to the range $0 \leq S \leq Gm^2/c$, and $m$ is the mass of the black
hole. As can be seen in Table \ref{tab:ul}, the spinning upper limits
are $\sim16\%$ larger than non-spinning. Signals from spinning systems 
are recovered with a worse match to our templates since we use a 
non-spinning template bank.

While the rates presented here represent an improvement over the
previously published results from earlier \ac{LIGO} and Virgo science
runs, they are still above the astrophysically predicted rates
of binary coalescence.  There are numerous uncertainties involved in
estimating astrophysical rates, including limited numbers of
observations and unknown model parameters; consequently the rate
estimates are rather uncertain.  For \ac{BNS} systems the estimated rates vary
between $1 \times 10^{-8}$ and $1 \times 10^{-5}$\,\perMpcyr, 
with a ``realistic'' estimate of $1 \times 10^{-6}$\,\perMpcyr. 
For \ac{BBH} and \ac{NSBH}, realistic estimates of the rate are
$5 \times 10^{-9}$\,\perMpcyr\ 
and $3 \times 10^{-8}$\,\perMpcyr\ 
with at least an order of
magnitude uncertainty in either direction \cite{ratesdoc}.  In all cases, the 
upper limits derived here are two to three orders of magnitude above the 
``realistic'' estimated rates, and about a factor of ten above the most optimistic
predictions.  These results are summarized in Figure
\ref{fig:rate_comp}.

\section{Discussion}
\label{sec:discussion}

We performed a search for gravitational waves from compact binary
coalescences with total mass between $2$ and $25\,\Msun$ with the
\ac{LIGO} and Virgo detectors using data taken between July 7, 2009 and
October 20, 2010.  No gravitational waves candidates were detected, and 
we placed new upper limits on \ac{CBC} rates.  These new limits are
up to a factor of 1.4 improvement over those achieved using 
previous LIGO and Virgo observational runs up to S5/VSR1 \cite{\sfivelvc}, 
but remain two to three orders of
magnitude above the astrophysically predicted rates.  

The installation of the advanced \ac{LIGO} and Virgo detectors has begun.
When operational, these detectors will provide a factor of ten increase
in sensitivity over the initial detectors, providing a factor of
$\sim1000$ increase in the sensitive volume.  At that time, we expect to
observe tens of binary coalescences per year \cite{ratesdoc}.

In order to detect this population of gravitational wave signals, we will have
to be able to confidently discriminate it from backgrounds caused by both
stationary and transient detector noise.  It is customary \cite{ratesdoc} to
assume that a signal with \ac{SNR} 8 in each detector would stand far enough
above background that we would consider it to be a detection candidate.  The
blind injection had somewhat larger \ac{SNR} than 8 in each detector, and we
were able estimate a \ac{FAR} of 1 in 7000 years for that event.
Alternatively, consider a coincident signal with exactly \ac{SNR} 8 in two
detectors.  Provided the signal is a good match to the template waveform
($\chi^{2}_{r} \approx 1$ in equation \ref{eq:new_snr}) this corresponds to
$\rho_c = 11.3$.  As can be seen from the extended background events with the
blind injection removed in Figure \ref{fig:far} (light-gray crosses), this
gives a \ac{FAR} of $\sim 1$ in $2 \times 10^{4}$ years in a single trial, or
$1$ in 3000 years over all trials.  Achieving similar-or-better background
distributions in Advanced LIGO and Virgo will require detailed data quality
studies of the detectors and feedback from the \ac{CBC} searches, along with
well-tuned signal-based vetoes. We have continued to develop the pipeline with
these goals in mind. For this analysis we significantly decreased the latency
between taking data and producing results, which allowed data quality vetoes to
be finely tuned for the \ac{CBC} search. These successes, along with the
successful recovery of the blind injection, give us confidence that we will be
able to detect gravitational waves from \ac{CBC}s at the expected rates in
Advanced LIGO and Virgo.

{\it Acknowledgments:}
The authors gratefully acknowledge the support of the United States
National Science Foundation for the construction and operation of the
LIGO Laboratory, the Science and Technology Facilities Council of the
United Kingdom, the Max-Planck-Society, and the State of
Niedersachsen/Germany for support of the construction and operation of
the GEO600 detector, and the Italian Istituto Nazionale di Fisica
Nucleare and the French Centre National de la Recherche Scientifique
for the construction and operation of the Virgo detector. The authors
also gratefully acknowledge the support of the research by these
agencies and by the Australian Research Council, 
the International Science Linkages program of the Commonwealth of Australia,
the Council of Scientific and Industrial Research of India, 
the Istituto Nazionale di Fisica Nucleare of Italy, 
the Spanish Ministerio de Educaci\'on y Ciencia, 
the Conselleria d'Economia Hisenda i Innovaci\'o of the
Govern de les Illes Balears, the Foundation for Fundamental Research
on Matter supported by the Netherlands Organisation for Scientific Research, 
the Polish Ministry of Science and Higher Education, the FOCUS
Programme of Foundation for Polish Science,
the Royal Society, the Scottish Funding Council, the
Scottish Universities Physics Alliance, The National Aeronautics and
Space Administration, the Carnegie Trust, the Leverhulme Trust, the
David and Lucile Packard Foundation, the Research Corporation, and
the Alfred P. Sloan Foundation.

\bibliography{../bibtex/iulpapers}

\begin{thebibliography}{38}
\expandafter\ifx\csname natexlab\endcsname\relax\def\natexlab#1{#1}\fi
\expandafter\ifx\csname bibnamefont\endcsname\relax
  \def\bibnamefont#1{#1}\fi
\expandafter\ifx\csname bibfnamefont\endcsname\relax
  \def\bibfnamefont#1{#1}\fi
\expandafter\ifx\csname citenamefont\endcsname\relax
  \def\citenamefont#1{#1}\fi
\expandafter\ifx\csname url\endcsname\relax
  \def\url#1{\texttt{#1}}\fi
\expandafter\ifx\csname urlprefix\endcsname\relax\def\urlprefix{URL }\fi
\providecommand{\bibinfo}[2]{#2}
\providecommand{\eprint}[2][]{\url{#2}}

\bibitem[{\citenamefont{Abbott et~al.}(2009{\natexlab{a}})}]{Abbott:2007kv}
\bibinfo{author}{\bibfnamefont{B.}~\bibnamefont{Abbott}} \bibnamefont{et~al.}
  (\bibinfo{collaboration}{LIGO Scientific Collaboration}),
  \bibinfo{journal}{Rept.~Prog.~Phys.} \textbf{\bibinfo{volume}{72}},
  \bibinfo{pages}{076901} (\bibinfo{year}{2009}{\natexlab{a}}),
  \eprint{arXiv:0711.3041}.

\bibitem[{\citenamefont{{Acernese} et~al.}(2008)}]{Acernese:2008b}
\bibinfo{author}{\bibfnamefont{F.}~\bibnamefont{{Acernese}}}
  \bibnamefont{et~al.}, \bibinfo{journal}{Class. Quant. Grav.}
  \textbf{\bibinfo{volume}{25}}, \bibinfo{pages}{184001}
  (\bibinfo{year}{2008}).

\bibitem[{GW1()}]{GW100916web}
\bibinfo{note}{LIGO Scientific Collaboration and Virgo Collaboration, The LIGO
  / Virgo Blind Injection GW100916 (2011)},
  \urlprefix\url{{http://www.ligo.org/science/GW100916/}}.

\bibitem[{\citenamefont{Abadie et~al.}(2010{\natexlab{a}})}]{S5LowMassLV}
\bibinfo{author}{\bibfnamefont{J.}~\bibnamefont{Abadie}} \bibnamefont{et~al.}
  (\bibinfo{collaboration}{LIGO Scientific Collaboration and Virgo
  Collaboration}), \bibinfo{journal}{\prd} \textbf{\bibinfo{volume}{82}},
  \bibinfo{pages}{102001} (\bibinfo{year}{2010}{\natexlab{a}}),
  \eprint{arXiv:1005.4655}.

\bibitem[{\citenamefont{Abadie et~al.}(2010{\natexlab{b}})}]{ratesdoc}
\bibinfo{author}{\bibfnamefont{J.}~\bibnamefont{Abadie}} \bibnamefont{et~al.}
  (\bibinfo{collaboration}{LIGO Scientific Collaboration and Virgo
  Collaboration}), \bibinfo{journal}{Class. Quant. Grav.}
  \textbf{\bibinfo{volume}{27}}, \bibinfo{pages}{173001}
  (\bibinfo{year}{2010}{\natexlab{b}}).

\bibitem[{\citenamefont{Adhikari et~al.}(2006)\citenamefont{Adhikari,
  Fritschel, and Waldman}}]{Adhikari:2006}
\bibinfo{author}{\bibfnamefont{R.}~\bibnamefont{Adhikari}},
  \bibinfo{author}{\bibfnamefont{P.}~\bibnamefont{Fritschel}},
  \bibnamefont{and} \bibinfo{author}{\bibfnamefont{S.}~\bibnamefont{Waldman}},
  \bibinfo{type}{Tech. Rep.} \bibinfo{number}{{LIGO}-T060156-v1},
  \bibinfo{institution}{{LIGO} Project} (\bibinfo{year}{2006}),
  \urlprefix\url{https://dcc.ligo.org/cgi-bin/DocDB/ShowDocument?docid=7384}.

\bibitem[{\citenamefont{Smith (for~the LIGO
  Scientific~Collaboration)}(2009)}]{Smith2009}
\bibinfo{author}{\bibfnamefont{J.}~\bibnamefont{Smith (for~the LIGO
  Scientific~Collaboration)}}, \bibinfo{journal}{Classical and Quantum Gravity}
  \textbf{\bibinfo{volume}{26}}, \bibinfo{pages}{114013}
  (\bibinfo{year}{2009}).

\bibitem[{\citenamefont{Fricke et~al.}(2011)}]{DCreadoutELIGO}
\bibinfo{author}{\bibfnamefont{T.}~\bibnamefont{Fricke}} \bibnamefont{et~al.}
  (\bibinfo{year}{2011}), \eprint{arXiv:1110:2815, submitted to
  Class.~Quant.~Grav.}

\bibitem[{\citenamefont{{Lorenzini (for the Virgo
  Collaboration)}}(2010)}]{Virgo:2010cqg}
\bibinfo{author}{\bibfnamefont{M.}~\bibnamefont{{Lorenzini (for the Virgo
  Collaboration)}}}, \bibinfo{journal}{Class. Quant. Grav.}
  \textbf{\bibinfo{volume}{27}}, \bibinfo{pages}{084021}
  (\bibinfo{year}{2010}).

\bibitem[{\citenamefont{Abbott
  et~al.}(2009{\natexlab{b}})}]{Collaboration:2009tt}
\bibinfo{author}{\bibfnamefont{B.}~\bibnamefont{Abbott}} \bibnamefont{et~al.}
  (\bibinfo{collaboration}{LIGO Scientific Collaboration}),
  \bibinfo{journal}{Phys.~Rev.~D} \textbf{\bibinfo{volume}{79}},
  \bibinfo{pages}{122001} (\bibinfo{year}{2009}{\natexlab{b}}),
  \eprint{arXiv:0901.0302}.

\bibitem[{\citenamefont{Abbott et~al.}(2009{\natexlab{c}})}]{Abbott:2009qj}
\bibinfo{author}{\bibfnamefont{B.}~\bibnamefont{Abbott}} \bibnamefont{et~al.}
  (\bibinfo{collaboration}{LIGO Scientific Collaboration}),
  \bibinfo{journal}{Phys.~Rev.~D} \textbf{\bibinfo{volume}{80}},
  \bibinfo{pages}{047101} (\bibinfo{year}{2009}{\natexlab{c}}).

\bibitem[{\citenamefont{Allen et~al.}(2012)\citenamefont{Allen, Anderson,
  Brady, Brown, and Creighton}}]{Allen:2005fk}
\bibinfo{author}{\bibfnamefont{B.}~\bibnamefont{Allen}},
  \bibinfo{author}{\bibfnamefont{W.~G.} \bibnamefont{Anderson}},
  \bibinfo{author}{\bibfnamefont{P.~R.} \bibnamefont{Brady}},
  \bibinfo{author}{\bibfnamefont{D.~A.} \bibnamefont{Brown}}, \bibnamefont{and}
  \bibinfo{author}{\bibfnamefont{J.~D.~E.} \bibnamefont{Creighton}}
  (\bibinfo{year}{2012}), \eprint{arXiv:gr-qc/0509116, submitted to
  Phys.~Rev.~D}.

\bibitem[{\citenamefont{Babak et~al.}(2006)\citenamefont{Babak,
  Balasubramanian, Churches, Cokelaer, and Sathyaprakash}}]{BBCCS:2006}
\bibinfo{author}{\bibfnamefont{S.}~\bibnamefont{Babak}},
  \bibinfo{author}{\bibfnamefont{R.}~\bibnamefont{Balasubramanian}},
  \bibinfo{author}{\bibfnamefont{D.}~\bibnamefont{Churches}},
  \bibinfo{author}{\bibfnamefont{T.}~\bibnamefont{Cokelaer}}, \bibnamefont{and}
  \bibinfo{author}{\bibfnamefont{B.~S.} \bibnamefont{Sathyaprakash}},
  \bibinfo{journal}{Class.\ Quant.\ Grav.} \textbf{\bibinfo{volume}{23}},
  \bibinfo{pages}{5477} (\bibinfo{year}{2006}), \eprint{gr-qc/0604037}.

\bibitem[{\citenamefont{Blanchet et~al.}(1995)\citenamefont{Blanchet, Damour,
  Iyer, Will, and Wiseman}}]{Blanchet:1995ez}
\bibinfo{author}{\bibfnamefont{L.}~\bibnamefont{Blanchet}},
  \bibinfo{author}{\bibfnamefont{T.}~\bibnamefont{Damour}},
  \bibinfo{author}{\bibfnamefont{B.~R.} \bibnamefont{Iyer}},
  \bibinfo{author}{\bibfnamefont{C.~M.} \bibnamefont{Will}}, \bibnamefont{and}
  \bibinfo{author}{\bibfnamefont{A.~G.} \bibnamefont{Wiseman}},
  \bibinfo{journal}{Phys. Rev. Lett.} \textbf{\bibinfo{volume}{74}},
  \bibinfo{pages}{3515} (\bibinfo{year}{1995}).

\bibitem[{\citenamefont{Blanchet et~al.}(2004)\citenamefont{Blanchet, Damour,
  Esposito-Far\`ese, and Iyer}}]{Blanchet:2004ek}
\bibinfo{author}{\bibfnamefont{L.}~\bibnamefont{Blanchet}},
  \bibinfo{author}{\bibfnamefont{T.}~\bibnamefont{Damour}},
  \bibinfo{author}{\bibfnamefont{G.}~\bibnamefont{Esposito-Far\`ese}},
  \bibnamefont{and} \bibinfo{author}{\bibfnamefont{B.~R.} \bibnamefont{Iyer}},
  \bibinfo{journal}{Phys. Rev. Lett.} \textbf{\bibinfo{volume}{93}},
  \bibinfo{pages}{091101} (\bibinfo{year}{2004}), \eprint{gr-qc/0406012}.

\bibitem[{\citenamefont{Abadie et~al.}(2011)}]{Collaboration:S5HighMass}
\bibinfo{author}{\bibfnamefont{J.}~\bibnamefont{Abadie}} \bibnamefont{et~al.}
  (\bibinfo{collaboration}{LIGO Scientific Collaboration and Virgo
  Collaboration}), \bibinfo{journal}{Phys. Rev. D}
  \textbf{\bibinfo{volume}{83}}, \bibinfo{pages}{122005}
  (\bibinfo{year}{2011}), \eprint{arXiv:1102.3781}.

\bibitem[{\citenamefont{Van Den~Broeck et~al.}(2009)}]{VanDenBroeck:2009gd}
\bibinfo{author}{\bibfnamefont{C.}~\bibnamefont{Van Den~Broeck}}
  \bibnamefont{et~al.}, \bibinfo{journal}{Phys.~Rev.~D}
  \textbf{\bibinfo{volume}{80}}, \bibinfo{pages}{024009}
  (\bibinfo{year}{2009}).

\bibitem[{\citenamefont{Robinson et~al.}(2008)\citenamefont{Robinson,
  Sathyaprakash, and Sengupta}}]{Robinson:2008}
\bibinfo{author}{\bibfnamefont{C.~A.~K.} \bibnamefont{Robinson}},
  \bibinfo{author}{\bibfnamefont{B.~S.} \bibnamefont{Sathyaprakash}},
  \bibnamefont{and} \bibinfo{author}{\bibfnamefont{A.~S.}
  \bibnamefont{Sengupta}}, \bibinfo{journal}{Phys.~Rev.~D}
  \textbf{\bibinfo{volume}{78}}, \bibinfo{eid}{062002} (\bibinfo{year}{2008}).

\bibitem[{\citenamefont{Allen}(2005)}]{Allen:2004}
\bibinfo{author}{\bibfnamefont{B.}~\bibnamefont{Allen}},
  \bibinfo{journal}{Phys.~Rev.~D} \textbf{\bibinfo{volume}{71}},
  \bibinfo{pages}{062001} (\bibinfo{year}{2005}).

\bibitem[{\citenamefont{Brown et~al.}(2012)}]{ihopePaper:2012}
\bibinfo{author}{\bibfnamefont{D.~A.} \bibnamefont{Brown}}
  \bibnamefont{et~al.}, \bibinfo{journal}{In preparation}
  (\bibinfo{year}{2012}).

\bibitem[{\citenamefont{{Christensen (for the LIGO Scientific Collaboration and
  the Virgo Collaboration)}}(2010)}]{Christensen:2010}
\bibinfo{author}{\bibfnamefont{N.}~\bibnamefont{{Christensen (for the LIGO
  Scientific Collaboration and the Virgo Collaboration)}}},
  \bibinfo{journal}{Class.\ Quantum Grav.} \textbf{\bibinfo{volume}{27}},
  \bibinfo{pages}{194010} (\bibinfo{year}{2010}).

\bibitem[{\citenamefont{{Robinet (for the LIGO Scientific Collaboration and the
  Virgo Collaboration)}}(2010)}]{Robinet:2010zz}
\bibinfo{author}{\bibfnamefont{F.}~\bibnamefont{{Robinet (for the LIGO
  Scientific Collaboration and the Virgo Collaboration)}}},
  \bibinfo{journal}{Class. Quant. Grav.} \textbf{\bibinfo{volume}{27}},
  \bibinfo{pages}{194012} (\bibinfo{year}{2010}).

\bibitem[{\citenamefont{Macleod et~al.}(2011)\citenamefont{Macleod, Fairhurst,
  Hughey, Lundgren, Pekowsky, Rollins, and Smith}}]{SeisVeto}
\bibinfo{author}{\bibfnamefont{D.~M.} \bibnamefont{Macleod}},
  \bibinfo{author}{\bibfnamefont{S.}~\bibnamefont{Fairhurst}},
  \bibinfo{author}{\bibfnamefont{B.}~\bibnamefont{Hughey}},
  \bibinfo{author}{\bibfnamefont{A.~P.} \bibnamefont{Lundgren}},
  \bibinfo{author}{\bibfnamefont{L.}~\bibnamefont{Pekowsky}},
  \bibinfo{author}{\bibfnamefont{J.}~\bibnamefont{Rollins}}, \bibnamefont{and}
  \bibinfo{author}{\bibfnamefont{J.~R.} \bibnamefont{Smith}}
  (\bibinfo{year}{2011}), \eprint{arXiv:1108:0312, submitted to Class. Quant.
  Grav.}

\bibitem[{\citenamefont{Abadie et~al.}(in preparation)}]{VirgoDetChar}
\bibinfo{author}{\bibfnamefont{J.}~\bibnamefont{Abadie}} \bibnamefont{et~al.}
  (\bibinfo{collaboration}{LIGO Scientific Collaboration and Virgo
  Collaboration}), \bibinfo{journal}{Class. Quant. Grav.}  (\bibinfo{year}{in
  preparation}).

\bibitem[{\citenamefont{Christensen}(1992)}]{Christensen:1992}
\bibinfo{author}{\bibfnamefont{N.}~\bibnamefont{Christensen}},
  \bibinfo{journal}{Phys. Rev. D} \textbf{\bibinfo{volume}{46}},
  \bibinfo{pages}{5250} (\bibinfo{year}{1992}).

\bibitem[{\citenamefont{{S}ingh and {R}{\"o}nnmark}(2004)}]{Singh:2004}
\bibinfo{author}{\bibfnamefont{A.~K.} \bibnamefont{{S}ingh}} \bibnamefont{and}
  \bibinfo{author}{\bibfnamefont{K.}~\bibnamefont{{R}{\"o}nnmark}},
  \bibinfo{journal}{{A}nnales {G}eophysicae} \textbf{\bibinfo{volume}{22}},
  \bibinfo{pages}{2067} (\bibinfo{year}{2004}).

\bibitem[{\citenamefont{{Veitch} and {Vecchio}}(2010)}]{Veitch:2010}
\bibinfo{author}{\bibfnamefont{J.}~\bibnamefont{{Veitch}}} \bibnamefont{and}
  \bibinfo{author}{\bibfnamefont{A.}~\bibnamefont{{Vecchio}}},
  \bibinfo{journal}{\prd} \textbf{\bibinfo{volume}{81}},
  \bibinfo{pages}{062003} (\bibinfo{year}{2010}), \eprint{0911.3820}.

\bibitem[{\citenamefont{{Feroz} et~al.}(2009)\citenamefont{{Feroz}, {Hobson},
  and {Bridges}}}]{Feroz:2009}
\bibinfo{author}{\bibfnamefont{F.}~\bibnamefont{{Feroz}}},
  \bibinfo{author}{\bibfnamefont{M.~P.} \bibnamefont{{Hobson}}},
  \bibnamefont{and}
  \bibinfo{author}{\bibfnamefont{M.}~\bibnamefont{{Bridges}}},
  \bibinfo{journal}{\mnras} \textbf{\bibinfo{volume}{398}},
  \bibinfo{pages}{1601} (\bibinfo{year}{2009}), \eprint{0809.3437}.

\bibitem[{\citenamefont{{van der Sluys}
  et~al.}(2008{\natexlab{a}})\citenamefont{{van der Sluys}, {Raymond},
  {Mandel}, {R{\"o}ver}, {Christensen}, {Kalogera}, {Meyer}, and
  {Vecchio}}}]{Sluys:2008a}
\bibinfo{author}{\bibfnamefont{M.}~\bibnamefont{{van der Sluys}}},
  \bibinfo{author}{\bibfnamefont{V.}~\bibnamefont{{Raymond}}},
  \bibinfo{author}{\bibfnamefont{I.}~\bibnamefont{{Mandel}}},
  \bibinfo{author}{\bibfnamefont{C.}~\bibnamefont{{R{\"o}ver}}},
  \bibinfo{author}{\bibfnamefont{N.}~\bibnamefont{{Christensen}}},
  \bibinfo{author}{\bibfnamefont{V.}~\bibnamefont{{Kalogera}}},
  \bibinfo{author}{\bibfnamefont{R.}~\bibnamefont{{Meyer}}}, \bibnamefont{and}
  \bibinfo{author}{\bibfnamefont{A.}~\bibnamefont{{Vecchio}}},
  \bibinfo{journal}{Classical and Quantum Gravity}
  \textbf{\bibinfo{volume}{25}}, \bibinfo{pages}{184011}
  (\bibinfo{year}{2008}{\natexlab{a}}), \eprint{0805.1689}.

\bibitem[{\citenamefont{{van der Sluys}
  et~al.}(2008{\natexlab{b}})\citenamefont{{van der Sluys}, {R{\"o}ver},
  {Stroeer}, {Raymond}, {Mandel}, {Christensen}, {Kalogera}, {Meyer}, and
  {Vecchio}}}]{Sluys:2008b}
\bibinfo{author}{\bibfnamefont{M.~V.} \bibnamefont{{van der Sluys}}},
  \bibinfo{author}{\bibfnamefont{C.}~\bibnamefont{{R{\"o}ver}}},
  \bibinfo{author}{\bibfnamefont{A.}~\bibnamefont{{Stroeer}}},
  \bibinfo{author}{\bibfnamefont{V.}~\bibnamefont{{Raymond}}},
  \bibinfo{author}{\bibfnamefont{I.}~\bibnamefont{{Mandel}}},
  \bibinfo{author}{\bibfnamefont{N.}~\bibnamefont{{Christensen}}},
  \bibinfo{author}{\bibfnamefont{V.}~\bibnamefont{{Kalogera}}},
  \bibinfo{author}{\bibfnamefont{R.}~\bibnamefont{{Meyer}}}, \bibnamefont{and}
  \bibinfo{author}{\bibfnamefont{A.}~\bibnamefont{{Vecchio}}},
  \bibinfo{journal}{\apjl} \textbf{\bibinfo{volume}{688}}, \bibinfo{pages}{L61}
  (\bibinfo{year}{2008}{\natexlab{b}}), \eprint{0710.1897}.

\bibitem[{\citenamefont{R\"{o}ver}(2007)}]{Roever2007}
\bibinfo{author}{\bibfnamefont{C.}~\bibnamefont{R\"{o}ver}}, Ph.D. thesis,
  \bibinfo{school}{The University of Auckland} (\bibinfo{year}{2007}),
  \bibinfo{note}{{URL} \url{http://hdl.handle.net/2292/2356}}.

\bibitem[{\citenamefont{Brady and Fairhurst}(2008)}]{Fairhurst:2007qj}
\bibinfo{author}{\bibfnamefont{P.~R.} \bibnamefont{Brady}} \bibnamefont{and}
  \bibinfo{author}{\bibfnamefont{S.}~\bibnamefont{Fairhurst}},
  \bibinfo{journal}{Class. Quantum Grav.} \textbf{\bibinfo{volume}{25}},
  \bibinfo{pages}{105002} (\bibinfo{year}{2008}), \eprint{arXiv:0707.2410}.

\bibitem[{\citenamefont{Brady et~al.}(2004)\citenamefont{Brady, Creighton, and
  Wiseman}}]{loudestGWDAW03}
\bibinfo{author}{\bibfnamefont{P.~R.} \bibnamefont{Brady}},
  \bibinfo{author}{\bibfnamefont{J.~D.~E.} \bibnamefont{Creighton}},
  \bibnamefont{and} \bibinfo{author}{\bibfnamefont{A.~G.}
  \bibnamefont{Wiseman}}, \bibinfo{journal}{Class. Quantum Grav.}
  \textbf{\bibinfo{volume}{21}}, \bibinfo{pages}{S1775} (\bibinfo{year}{2004}).

\bibitem[{\citenamefont{Biswas et~al.}(2009)\citenamefont{Biswas, Brady,
  Creighton, and Fairhurst}}]{Biswas:2007ni}
\bibinfo{author}{\bibfnamefont{R.}~\bibnamefont{Biswas}},
  \bibinfo{author}{\bibfnamefont{P.~R.} \bibnamefont{Brady}},
  \bibinfo{author}{\bibfnamefont{J.~D.~E.} \bibnamefont{Creighton}},
  \bibnamefont{and}
  \bibinfo{author}{\bibfnamefont{S.}~\bibnamefont{Fairhurst}},
  \bibinfo{journal}{Class. Quantum Grav.} \textbf{\bibinfo{volume}{26}},
  \bibinfo{pages}{175009} (\bibinfo{year}{2009}), \eprint{arXiv:0710.0465}.

\bibitem[{\citenamefont{Kopparapu et~al.}(2008)\citenamefont{Kopparapu, Hanna,
  Kalogera, O'Shaughnessy, Gonzalez, Brady, and Fairhurst}}]{LIGOS3S4Galaxies}
\bibinfo{author}{\bibfnamefont{R.~K.} \bibnamefont{Kopparapu}},
  \bibinfo{author}{\bibfnamefont{C.}~\bibnamefont{Hanna}},
  \bibinfo{author}{\bibfnamefont{V.}~\bibnamefont{Kalogera}},
  \bibinfo{author}{\bibfnamefont{R.}~\bibnamefont{O'Shaughnessy}},
  \bibinfo{author}{\bibfnamefont{G.}~\bibnamefont{Gonzalez}},
  \bibinfo{author}{\bibfnamefont{P.~R.} \bibnamefont{Brady}}, \bibnamefont{and}
  \bibinfo{author}{\bibfnamefont{S.}~\bibnamefont{Fairhurst}},
  \bibinfo{journal}{\apj} \textbf{\bibinfo{volume}{675}}, \bibinfo{pages}{1459}
  (\bibinfo{year}{2008}).

\bibitem[{\citenamefont{{Finn} and {Chernoff}}(1993)}]{FinnChernoff:1993}
\bibinfo{author}{\bibfnamefont{L.~S.} \bibnamefont{{Finn}}} \bibnamefont{and}
  \bibinfo{author}{\bibfnamefont{D.~F.} \bibnamefont{{Chernoff}}},
  \bibinfo{journal}{\prd} \textbf{\bibinfo{volume}{47}}, \bibinfo{pages}{2198}
  (\bibinfo{year}{1993}), \eprint{arXiv:gr-qc/9301003}.

\bibitem[{\citenamefont{Manchester et~al.}(2005)\citenamefont{Manchester,
  Hobbs, Teoh, and Hobbs}}]{ATNF:psrcat}
\bibinfo{author}{\bibfnamefont{R.~N.} \bibnamefont{Manchester}},
  \bibinfo{author}{\bibfnamefont{G.~B.} \bibnamefont{Hobbs}},
  \bibinfo{author}{\bibfnamefont{A.}~\bibnamefont{Teoh}}, \bibnamefont{and}
  \bibinfo{author}{\bibfnamefont{M.}~\bibnamefont{Hobbs}},
  \bibinfo{journal}{Astronom. J.} \textbf{\bibinfo{volume}{129}},
  \bibinfo{pages}{1993} (\bibinfo{year}{2005}).

\bibitem[{\citenamefont{Apostolatos et~al.}(1994)\citenamefont{Apostolatos,
  Cutler, Sussman, and Thorne}}]{Apostolatos:1994}
\bibinfo{author}{\bibfnamefont{T.~A.} \bibnamefont{Apostolatos}},
  \bibinfo{author}{\bibfnamefont{C.}~\bibnamefont{Cutler}},
  \bibinfo{author}{\bibfnamefont{G.~J.} \bibnamefont{Sussman}},
  \bibnamefont{and} \bibinfo{author}{\bibfnamefont{K.~S.}
  \bibnamefont{Thorne}}, \bibinfo{journal}{Phys.~Rev.~D}
  \textbf{\bibinfo{volume}{49}}, \bibinfo{pages}{6274} (\bibinfo{year}{1994}).

\end{thebibliography}

\end{document}